\documentclass[%
reprint,
amsmath,amssymb,
aps,
prd,
superscriptaddress,
]{revtex4-2}

\usepackage{ORCIDinREVTeX}
\usepackage{graphicx}
\usepackage{dcolumn}
\usepackage[caption=false]{subfig}
\usepackage{lmodern} 
\usepackage{courier} 
\usepackage[section]{placeins}
\usepackage{natbib}
\usepackage{bm}
\usepackage{hyperref}
\hypersetup{colorlinks=true, citecolor=cyan, urlcolor = blue, linkcolor = blue}

\usepackage{aas_macros}
\usepackage{color}

\DeclareUnicodeCharacter{2009}{\,}

\begin{document}

\title{The NANOGrav 15-year Data Set: Search for Gravitational Wave Memory}
\author{Gabriella Agazie}
\orcid{0000-0001-5134-3925}
\affiliation{Center for Gravitation, Cosmology and Astrophysics, Department of Physics, University of Wisconsin-Milwaukee,\\ P.O. Box 413, Milwaukee, WI 53201, USA}
\author{Akash Anumarlapudi}
\orcid{0000-0002-8935-9882}
\affiliation{Center for Gravitation, Cosmology and Astrophysics, Department of Physics, University of Wisconsin-Milwaukee,\\ P.O. Box 413, Milwaukee, WI 53201, USA}
\author{Anne M. Archibald}
\orcid{0000-0003-0638-3340}
\affiliation{Newcastle University, NE1 7RU, UK}
\author{Zaven Arzoumanian}
\affiliation{X-Ray Astrophysics Laboratory, NASA Goddard Space Flight Center, Code 662, Greenbelt, MD 20771, USA}
\author{Jeremy G. Baier}
\orcid{0000-0002-4972-1525}
\affiliation{Department of Physics, Oregon State University, Corvallis, OR 97331, USA}
\author{Paul T. Baker}
\orcid{0000-0003-2745-753X}
\affiliation{Department of Physics and Astronomy, Widener University, One University Place, Chester, PA 19013, USA}
\author{Bence B\'{e}csy}
\orcid{0000-0003-0909-5563}
\affiliation{Department of Physics, Oregon State University, Corvallis, OR 97331, USA}
\author{Laura Blecha}
\orcid{0000-0002-2183-1087}
\affiliation{Physics Department, University of Florida, Gainesville, FL 32611, USA}
\author{Adam Brazier}
\orcid{0000-0001-6341-7178}
\affiliation{Cornell Center for Astrophysics and Planetary Science and Department of Astronomy, Cornell University, Ithaca, NY 14853, USA}
\affiliation{Cornell Center for Advanced Computing, Cornell University, Ithaca, NY 14853, USA}
\author{Paul R. Brook}
\orcid{0000-0003-3053-6538}
\affiliation{Institute for Gravitational Wave Astronomy and School of Physics and Astronomy, University of Birmingham, Edgbaston, Birmingham B15 2TT, UK}
% ...

%


\date{\today}

\begin{abstract}
We present the results of a search for nonlinear gravitational wave memory in the NANOGrav 15-year data set. We find no significant evidence for memory signals in the dataset, with a maximum Bayes factor of $3.1$ in favor of a model including memory. We therefore place upper limits on the strain of potential gravitational wave memory events as a function of sky location and observing epoch. We find upper limits that are not always more constraining than previous NANOGrav results. We show that it is likely due to the increase in common red noise between the $12.5$-year and $15$-year NANOGrav datasets.
\end{abstract}

\maketitle


\section{\label{sec:intro} Introduction} 

Recent searches for gravitational waves (GWs) in Pulsar Timing Array (PTA) data have yielded strong evidence for the presence of a gravitational wave background (GWB) \cite{agazie_nanograv_2023-1, agazie_nanograv_2023-3, xu_searching_2023, PPTA_DR3_data, ppta_dr3_gwb, EPTA_DR2_data, epta_collaboration_and_inpta_collaboration_second_2023}. With this new discovery, we are able to start probing the stochastic GW background produced in the nanohertz band \cite{agazie_nanograv_2023, NG15_NoiseBudget, nanograv_15yr_newphysics, Agazie_BHBconstraints_2023, Anisotropy_2023, Afzal_NewPhysics_2023}, as well as other types of GW signals, including continuous waves \cite{ellis_optimal_2012, falxa_searching_2023, arzoumanian_nanograv_2023, antoniadis_second_2024}, bursts with memory \cite{12p5yr_NG_BWM, madison_pulsar_2017, sun_implementation_2023}, and generic bursts \cite{becsy_bayesian_2021, deng_searching_2023, taylor2024fastwaveletbasissearch}. In this paper, we present the results of our searches for nonlinear gravitational wave memory in the NANOGrav 15-year data set \citep{agazie_nanograv_2023-3}. 

A PTA is a collection of millisecond pulsars (MSPs) with ultra-stable rotational periods \cite{lorimer_binary_2008}. We can measure the times of arrival (TOAs) of the radio pulses from some of these MSPs to a precision of a few tens of nanoseconds for our best-timed pulsars. The pulse TOAs from an MSP depend on its period, spin-down rate, position, and motion across the sky, as well as orbital motion if the MSP is in a binary. Models that account for these effects can be used to predict the time at which each pulse is expected to arrive, and ``timing residuals" are calculated by subtracting the model-predicted TOAs from the measured TOAs. GWs cause fluctuations in the TOAs that are correlated among MSPs \cite{sazhin_opportunities_1978, detweiler_pulsar_1979, hellings_upper_1983}. 

Nonlinear GW memory is a constant shift in the gravitational metric due to violent GW-emitting events, such as mergers of super massive black hole binaries (SMBHBs) \cite{thorne_gravitational-wave_1992, christodoulou_nonlinear_1991, wiseman_christodoulous_1991,islo_prospects_2019}. As a memory wavefront passes the Earth or the pulsar, its effect is to redshift or blueshift the train of pulses traveling towards Earth. This appears to us as a sudden change in the rotational frequency of the pulsar, producing a linear drift in our timing residuals which we will refer to as a ramp. 

There have been several searches for GW memory in NANOGrav datasets \cite{12p5yr_NG_BWM, aggarwal_nanograv_2020, arzoumanian_nanograv_2015}, which have resulted in upper limits on potential strains and event rates of memory events. There are also searches planned and ongoing for other ground and space based detectors like LISA and LIGO \cite{favata_nonlinear_2009, amaro-seoane_laser_2017, boersma_forecasts_2020}.

Here we report the results of a search for GW memory in the latest NANOGrav $15$-year dataset \cite{agazie_nanograv_2023-3}. We find no compelling evidence for a GW memory event in this dataset. A Bayesian model selection analysis only yielded a Bayes factor of $\mathrm{BF} = 3.1$ in favor of a model including a GW memory event in the last three years of the data set. Furthermore, GW memory bursts which occur late in a PTA data set are not credible since there are too few TOAs present after the event to convincingly rule out a noise transient. We also show that ``hot spots" present in previous analyses are less significant in this dataset \cite{12p5yr_NG_BWM}.

Having found no significant evidence for GW memory, we place updated upper limits on GW memory strain and event rates. Following the procedure in used in \cite{12p5yr_NG_BWM}, we use single-pulsar likelihood lookup tables to expedite the calculation of global likelihoods \cite{sun_implementation_2023}. These upper limits do not uniformly yield better constraints over the $12.5$-year dataset results \cite{12p5yr_NG_BWM}. The latter are likely the effect of the increased common red noise signal we found in previous GW searches \cite{agazie_nanograv_2023}.

We begin \autoref{sec:Signal} by reviewing the signal model for a GW memory event in PTA data. In \autoref{sec:Methods} we detail the methods used to perform Bayesian model selection and compute upper limits on the GW memory strain. Following that, in \autoref{sec:Results} we present the results of those searches and upper limit calculations. Finally, we present some conclusions in \autoref{sec:Conclusion}.

\section{\label{sec:Signal} Signal} 

GW memory is a permanent change in the space-time metric resulting from the momentum carried away from a system by GWs. In the context of PTA experiments, which measure the TOAs of radio pulses from pulsars, when a GW memory wavefront passes over the Earth, we observe an apparent shift in the rotational frequencies of all pulsars in the data set. The sudden frequency change results in a constant excess or deficit of phase with each rotation of the pulsar, which in turn produces a ``ramp"-shaped signature in the timing residuals.


We can write the residuals of a pulsar in the direction $\hat{\mathbf{p}}$ induced by a GW memory wave front passing Earth in the direction $\hat{\mathbf{k}}$ with a polarization $\psi$ as 
\begin{equation}\label{eq:t_mem}
    \delta \mathbf{t}_{mem}(t) =  B(\hat{\mathbf{k}},\hat{\mathbf{p}},\psi) h_{mem}(t).
\end{equation}
Here, $B(\hat{\mathbf{k}},\hat{\mathbf{p}},\psi)$ accounts for the response of the Earth-pulsar system to a GW 
\begin{equation}\label{eq:B_factor}    B(\hat{\mathbf{k}},\hat{\mathbf{p}},\psi) = \frac{1}{2}\cos{(2\psi)}(1-\cos{\alpha}),
\end{equation}
where $\alpha$ is the angle between $\hat{\mathbf{p}}$ and $\hat{\mathbf{k}}$. The time dependent strain, $h_{mem}(t)$, is given by 
\begin{equation}\label{eq:h_mem}
    h_{mem}(t) = h_0[(t-t_0)\Theta(t-t_0) - (t-t_i)\Theta(t-t_i)],
\end{equation}
where $t_0$ is the time at which the memory wavefront passes by the Earth, and $t_i = t_0 + (\lvert\vec{p}\rvert/c)[1+\cos{\alpha}]$ is the time at which the wavefront passes by the $i^{th}$ pulsar, and $h_0$ is the strain of the GW memory signal.

In the equation above, we often call the ramp term with $t_0$ the ``Earth term," and the term with $t_i$ the ``Pulsar term." The ``Earth term" induces a ramp in the timing residuals when the GW wavefront passes over the Earth, and results in a change across all the pulsars. If a memory wavefront pass over a pulsar, we get a ``Pulsar term" event. This causes that individual pulsar's apparent rotational frequency to change. However, because the change only occurs in one pulsar, and the time it takes that wavefront to reach another pulsar vastly exceeds our dataset length, it is difficult to distinguish a single pulsar GW memory event from some other transient effect, such as a pulsar glitch or noise transient \cite{van_haasteren_gravitational-wave_2010, cordes_detecting_2012}. We therefore focus our analysis on the Earth term part of the signal.

The full timing residuals for a single pulsar in our PTA are
\begin{equation}\label{eq:residuals}
    \delta \mathbf{t} =  \delta \mathbf{t}_{mem} + M \boldsymbol{\epsilon} + \mathbf{n} + F\mathbf{a} + F_{gw}\mathbf{a}_{gw}.
\end{equation}
In this equation, $M$ is the design matrix which accounts for small errors in the pulsar timing model $\boldsymbol{\epsilon}$, and $\textbf{n}$ are the white noise uncertainties. The pulsar intrinsic red noise is modeled with a Fourier basis $F$ and Fourier coefficients $\textbf{a}$, while the common spatially uncorrelated red noise (CURN) process is modeled by the basis $F_{GW}$ and coefficients $\textbf{a}_{GW}$. 

We include both intrinsic pulsar red noise and a common red noise since both have been shown to be present in these PTA data sets \cite{agazie_nanograv_2023-1, arzoumanian_nanograv_2020}, and model the red noise processes using power laws \cite{phinney_practical_2001}
\begin{equation}\label{eq:red_noise_power}
    P_{RN}(f) =  A_{RN}^2 \left( \frac{f}{f_{1yr}} \right)^{-\gamma}.
\end{equation}
Here, $A_{RN}$ is the red noise amplitude at the reference frequency $f_{1yr}$, and $\gamma$ is the spectral index.

Although the 15-year data search has also revealed significant evidence that the common red noise process is due to GWs with Hellings-Downs correlations \cite{agazie_nanograv_2023-1, hellings_upper_1983}, we model it here as CURN because the methods used to make these analyses computationally efficient use a per-pulsar factorized likelihood (see \autoref{sec:Methods} for details). Including inter-pulsar correlations in our model would require several orders of magnitude more computational resources to handle, and modeling the common process as CURN still accounts for the majority of the effects of red noise.

\section{\label{sec:Methods} Methods} 

The NANOGrav 15-year dataset consists of the pulse times of arrival (TOAs) and timing models for $68$ pulsars over a 15 year timing baseline between July 2004 and August 2020. For this paper, we exclude pulsar J0614-3329, which has a timing baseline shorter than 3 years. A detailed description of the the dataset can be found in \cite{agazie_nanograv_2023-3}. 






To search for Earth-term GW memory, we use \texttt{enterprise} \cite{ellis_justin_a_enterprise_2020} and \texttt{enterprise extensions} \cite{Taylor_enterprise_extensions_2021} to perform model selection using Bayesian MCMC product-space sampling \cite{carlin_bayesian_1995, godsill_relationship_2001}. In these searches we use the likelihood as defined in section $3$ of \cite{arzoumanian_nanograv_2016}, as well as log uniform priors on the pulsar-intrinsic red noise amplitude, CURN amplitude, and GW memory amplitude. For the GW memory epoch, we use uniform priors which exclude early and late times. Prior work in Ref. \cite{sun_implementation_2023} shows that inclusion of early and late times can result in biased marginal posterior probability distributions of the GW strain amplitude. 

To compute upper limits, we begin by generating single pulsar likelihood lookup tables. These lookup tables contain the likelihood for the GW strain and marginalized over the pulsar-intrinsic red noise and CURN parameters. These can then be used to calculate the global likelihood because, as long as our noise covariance matrix is block diagonal, the likelihood can be factorized likelihood over pulsars. Specifically,
\begin{equation}\label{eq:likelihood}
    p(\delta \mathbf{t}|\hat{\mathbf{k}}, \psi, t_0, h_0) = \prod_{i=1}^{N_{psr}}  p_i(\delta \mathbf{t}|h_i, t_0),
\end{equation}
where $h_i = B(\hat{\mathbf{k}},\hat{\mathbf{p}}_i,\psi) \times h_0$. This improves the efficiency of calculations of the global Earth-term likelihoods by avoiding repeated expensive matrix inversions (see ref. \cite{sun_implementation_2023} for more details) 

When we numerically marginalize over the intrinsic red noise $\{A_{\mathrm{IRN}}, \gamma_{\mathrm{IRN}}\}$ and CURN $\{A_{\mathrm{CURN}}, \gamma_{\mathrm{CURN}}\}$ parameters for each pulsar, we use a log-uniform prior on the intrinsic red noise amplitude $\log_{10}{A_{\mathrm{IRN}}} \sim \mathrm{Uniform}(-16, -14)$ and a uniform prior on the spectral index $\gamma_{\mathrm{IRN}} \sim \mathrm{Uniform}(0, 7)$. For the CURN, we use the same prior for the amplitude as for the intrinsic red noise $\log_{10}{A_{\mathrm{CURN}}} \sim \mathrm{Uniform}(-16, -14)$. Rather than marginalizing over the spectral index, we fix it to the astrophysically predicted value for a background of GWs produced by SMBHBs $\gamma_{\mathrm{CURN}} = 13/3$. This generates a noise-marginalized likelihood lookup table, with the only remaining non-sky location parameters being \{$h_0, t_B,  \mathrm{sign} (B(\hat{\mathbf{k}},\hat{\mathbf{p}},\psi))$\}. 

\section{\label{sec:Results} Results} 

In this section, we present the results of Bayesian searches for GW memory events, and upper limits on the gravitational memory strain present in the NANOGrav 15-year PTA. 

\subsection{Signal Search}\label{ssec:SignalSearch} 

\begin{figure*}[htbp]
\subfloat[]{%
\includegraphics[width =1.0\columnwidth]{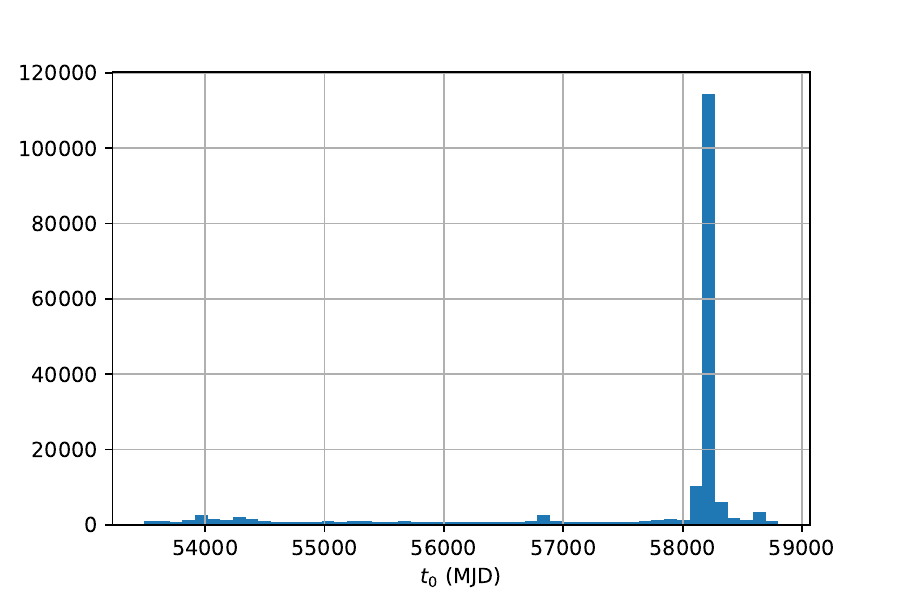}%
}\quad
\subfloat[]{
\includegraphics[width =1.0\columnwidth]{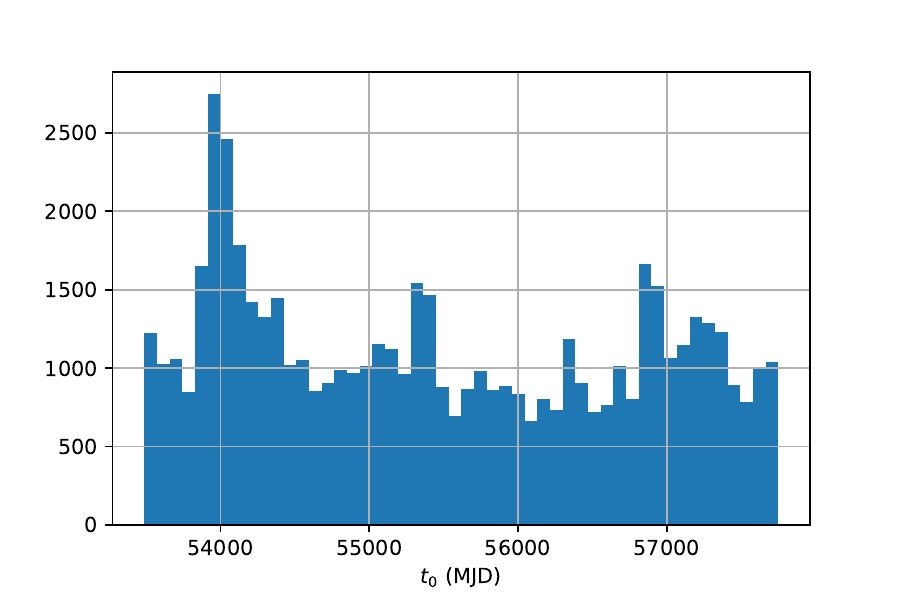}%
}\quad
\caption{These plots show the epoch ($t_0$) posterior from model selection analysis, which have been masked to include samples only from the model that includes intrinsic noise, a CURN process, and a GW memory signal. In \textbf{(a)} the prior on $t_0$ is allowed to be the length of the dataset with $270$ days removed from either end to ignore edge effects. We see that there is one discernible event at $58200$ MJD, however this model is only favored by a Bayes factor of around 3 compared to the model without a GW memory event. In \textbf{(b)} the prior is constrained to the length of the epochs present in the NANOGrav $12.5$-year search \cite{12p5yr_NG_BWM}. This provides more samples in the higher sensitivity central times of the dataset. The latter shows only a small feature at the time of the sampling gap around $54000$ MJD \cite{12p5yr_NG_BWM}. }\label{fig:t0_posteriors}
\end{figure*}

We start with the Bayesian model selection analyses, where the two models being compared are 1) a model with Gaussian white noise, per-pulsar red noise, CURN, and GW memory to 2) a model that includes only white noise, per-pulsar red noise, and CURN. We performed two analyses with slightly different priors on the burst epoch. The first prior excludes the first and last $270$ days to avoid spurious GW events at the edges of our dataset~\cite{12p5yr_NG_BWM}. Events that are found early in the data set are nearly impossible to rule out because they are covariant with the linear part of the quadratic fit. Events that are found very late in the data set have very few TOAs which support the ``ramp"-shaped signature. 
The second prior constrains the burst epoch to the same time baseline as the NANOGrav $12.5$-year data set for comparisons with previous results. 

The first analysis, that excludes $270$ days from both ends of the data set, yields a Bayes factor of $3.1$ with a standard deviation of $0.3$ in favor of including a GW memory burst around $t_0 = 58000$. As shown in \autoref{fig:t0_posteriors}(a), this is the only significant potential event in the epoch posterior. Since this event occurs in the last three years of the dataset, where we have low sensitivity due to the lack of data to constrain the ramp, and the Bayes factor is not significant, there is no compelling evidence for GW memory event.

The second analysis, which uses the same prior on burst epoch as the NANOGrav $12.5$-year data set (shown in  \autoref{fig:t0_posteriors}(b)) yields no evidence of any signals. This model selection run gave a Bayes factor of $1.22$ with a standard deviation of $0.04$ disfavoring the GW memory burst. The only visible peak, around $54000$ MJD, occurs in a well known data gap near the beginning of the dataset that has been seen in the $11$ year and $12.5$ year searches \cite{12p5yr_NG_BWM, aggarwal_nanograv_2020}. It is hard to constrain such signals because of the quadratic subtraction done when we fit the timing model to the pulsar TOAs. This removes some of the significance from bursts at early times, and leaves the most significant feature as a cusp appearing near the start of the ramp, which in this case sits inside of the aforementioned data gap.


\subsection{\label{sec:Limits} Upper Limits} 

Having found no significant evidence of a detection, we move on to present upper limits on GW memory events. We begin with strain amplitude upper limits vs. burst epoch, which inform constraints on rates of GW memory events. Then we look at strain amplitude upper limits vs. sky location. We compare all of these upper limits to those computed in the NANOGrav 12.5-year search for GW memory \citep{12p5yr_NG_BWM}. Finally, we also provide some context for the trends we see with regards to the red noise levels in the datasets.

\begin{figure}[htbp]
    \centering
    \includegraphics[width = 1\columnwidth]{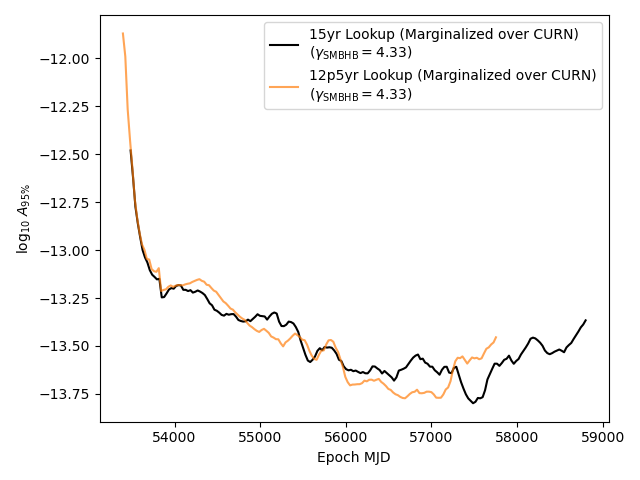}
    \caption{Upper limits on the GW memory strain amplitude ($h_0$) for bursts at various epochs. Comparing the 15-year and 12.5-year upper limits from \cite{12p5yr_NG_BWM}, we see improvements on the ends of the epoch time span, and differences at central times due to the time baseline for the two timing models, and the red noise levels in the two datasets.}
    \label{fig:epoch_UL}
\end{figure}

\begin{figure*}[htbp]
    \centering
    \includegraphics[width = 2\columnwidth]{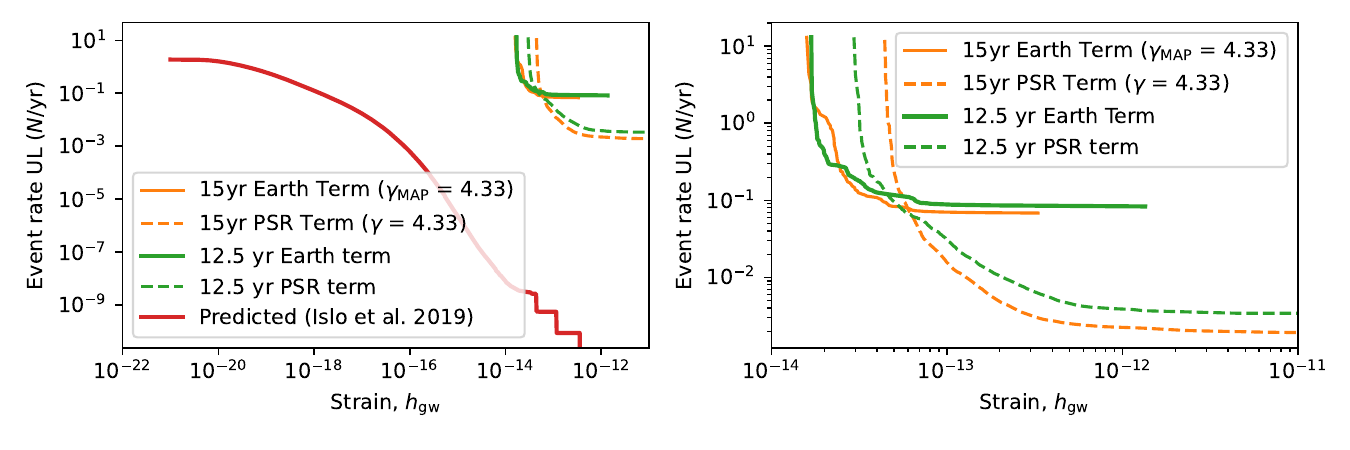}
    \caption{Upper limits on the GW memory event rate in the combined individual pulsar terms, as well as the earth term. These are presented vs. GW memory strain ($h_0$), with the (\textit{right}) plot presenting the same data zoomed in to a strain range of $10^{-14}$ to $10^{-11}$.}
    \label{fig:rate_UL}
\end{figure*}

\subsubsection{UL vs. Epoch}\label{ssec:epoch} 

\autoref{fig:epoch_UL} shows the strain upper limit as a function of burst epoch marginalized over sky location and polarization. The burst epoch is sampled using a uniform prior, while the sky location and polarization are marginalized by taking an equal number of samples from every source orientation bin to avoid bias from the least sensitive sky locations. We also use the burst epoch upper limits to calculate upper limits on the merger rates of SMBHBs in \autoref{fig:rate_UL}. We present these new upper limits and compare them with our previous $12.5$ year dataset results.

The 15-year upper limits deviate from the $12.5$-year upper limits in a number of ways. Starting with the strain upper limits, there is some improvement at early epochs in the $15$-year results. However, near the middle of the data set, there is a loss of sensitivity before our highest sensitivity around $57600$ MJD. We believe this to be the result of the quadratic subtraction necessary for our timing model, which reduces sensitivity near the center of each pulsars' timing baselines \cite{cordes_detecting_2012, van_haasteren_gravitational-wave_2010}. Now that we have more pulsars with data in the most recent 3 years, the sensitivity loss from quadratic subtraction that occurs in the middle of a dataset has shifted to a later period in the dataset. The small changes in our recovered CURN parameters also have a significant effect on the recovered upper limits. See sub-section \ref{ssec:scaling} for further discussion of this effect.

The event rate upper limits also show some differences from the $12.5$-year dataset. At high strains, we see more constrained event rate upper limits for both the Earth and pulsar terms, while at lower strains the event rate are less constrained when considering the pulsar term. These low strain event limits are dominated by our longest-timed pulsars, which are affected the most significantly by the change in red noise amplitude between datasets.





\subsubsection{UL vs. Sky Location}\label{ssec:skyLoc} 

\begin{figure}[htbp] 
    \includegraphics[width =1.0\columnwidth]{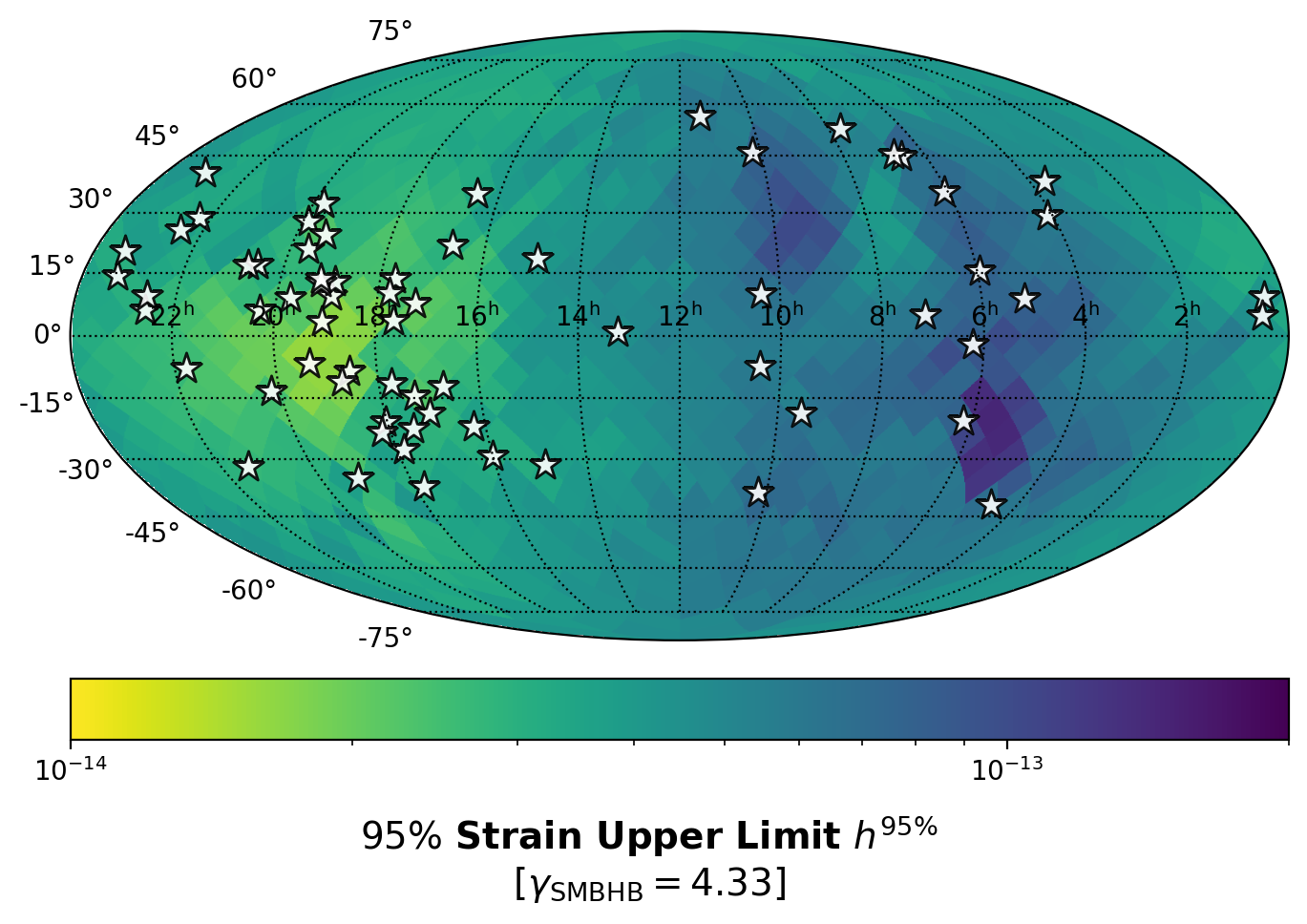}
    \includegraphics[width =1.0\columnwidth]{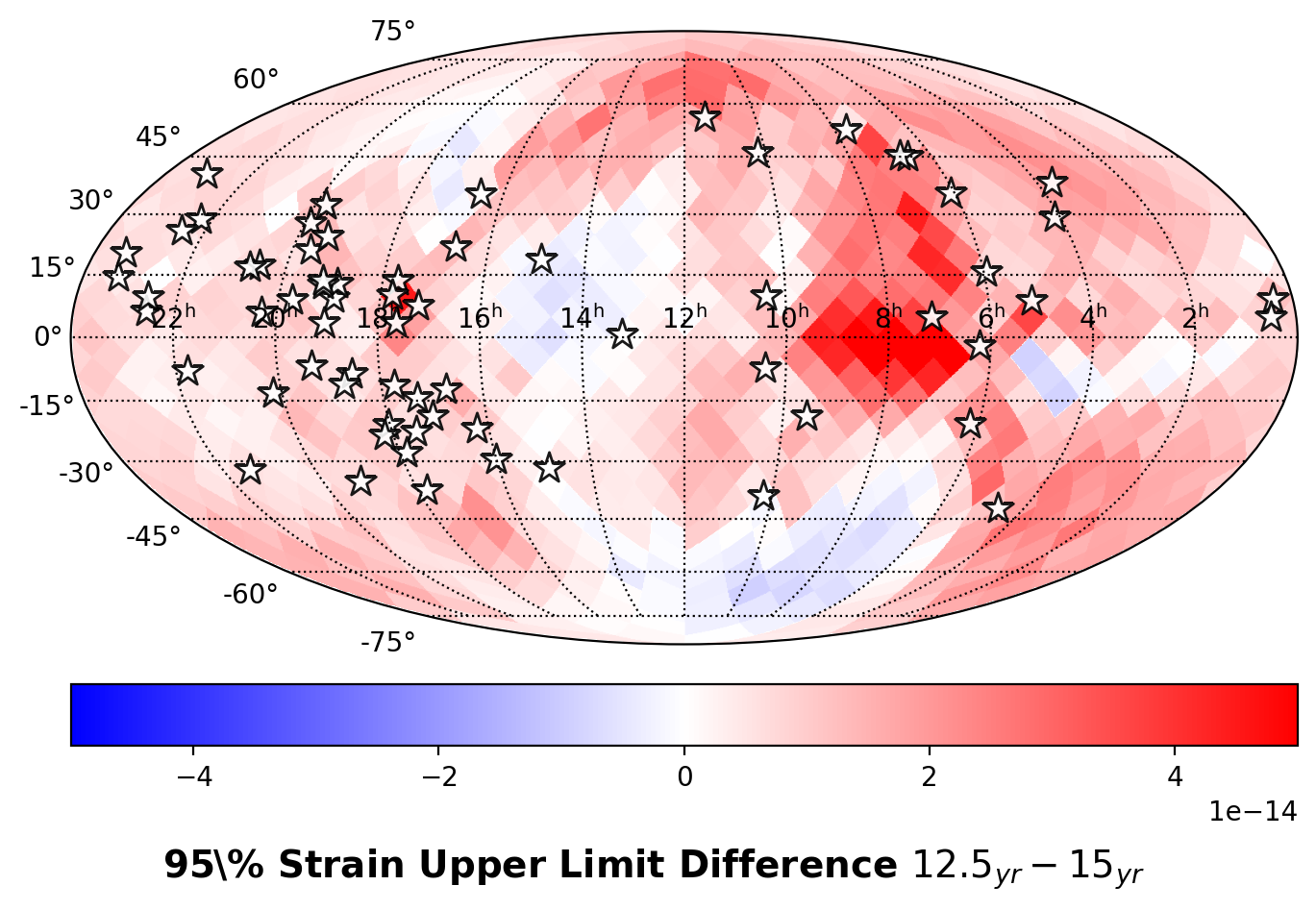}
    \caption{The (\textit{top}) plot is a upper limit on strain amplitude across the sky, averaged over the final 3 years of 15-Year NANOGrav dataset. The (\textit{bottom}) plot is the difference in upper limits between the 15-year and 12.5-year searches during the last 3 years of the $12.5$-year dataset.} 
    \label{fig:Sky_UL}
\end{figure}

We also present the epoch-marginalized upper limits as a function of source sky location. These upper limits show how sensitivity varies across the sky. To compute these upper limits, we marginalize over a specified range of GW memory epochs, while sampling over a grid of sky locations. This is displayed in the top plot of \autoref{fig:Sky_UL}. As expected, when using the last 3 years of data from the 15-year data set, we see the lowest upper limits in the most densely populated regions of the sky. However, the upper limits still do not reach below $h_0 = 10^{-14}$ in these regions.

 Figure \ref{fig:Sky_UL} shows the difference in upper limits on GW strain between the $15$-year the $12.5$-year analyses. Both are plots averaged over the last 3 year period of the 12-year dataset so that results are directly comparable. We see that over most of the sky there is an improvement in these upper limits in the $15$ year search. In other words, the $15$-year upper limits are lower in most parts of the sky. This is consistent with the differences we see over these times in the upper limit vs. epoch results above, as epochs near the end of our dataset have decreased upper limits for the $15$-year compared to the $12.5$-year results. In some of that region it is higher for the $15$-year, which is still consistent as the differences between the two results are not large.

\subsubsection{SNR Scaling Expectation}\label{ssec:scaling} 

Some of the above results are surprising, most notably the loss of sensitivity between MJDs of 56000 and 57000 seen in \autoref{fig:epoch_UL}. In order to better understand how red noise may affect GW memory sensitivity, we also performed a simple test of how GW memory signal to noise ratio (SNR) changes between the $12.5$-year and $15$-year memory searches. This test estimates how much of an increase in GW memory event detectability we expect to see given the extensions of our dataset, as well as the shifts we have seen in the amplitude of the GW background between these two datasets. To do this, we used a simple simulated dataset which includes a Gaussian white noise process, CURN, and GW memory event which starts $t_B = 4$ years into the data set with a strain of $h_0 = 10^{-14.5}$. The white noise level is set to $0.5$ microseconds for all pulsars. We fix the spectral index of the CURN to $\gamma=\frac{13}{3}$  and compare the resulting time evolution of the SNR while fixing the amplitude of the CURN to the maximum a posteriori values from the results of the NANOGrav $15$-year ($A=2.4\times10^{-15}$) and $12.5$-year ($A=1.9\times10^{-15}$) searches for a stochastic GW background. We calculate these SNRs by taking the inner product of the post timing model fit GW memory signal with itself using the R matrix formalism and the noise covariance matrix as described in \cite{Chamberlin_2015}. 

\begin{figure}[htbp]
    \centering
    \includegraphics[width = \columnwidth]{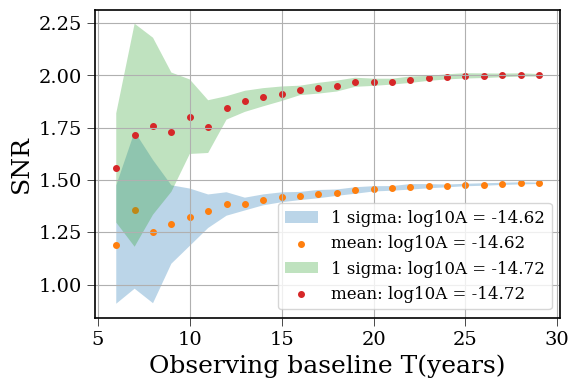}
    \caption{SNR of a GW memory signal given two potential CURN amplitudes, and over a variety of pulsar timing baselines. The green section with red dots has a CURN consistent with $12.5$-year dataset, while the blue section with orange dots has a CURN consistent with the $15$-year Both the mean and one sigma SNRs are calculated over 100 realizations of the noise.}
    \label{fig:SNR_comp}
\end{figure}

The results are shown in \autoref{fig:SNR_comp}, where we display the mean SNR and $1-\sigma$ uncertainty in the SNR for $100$ realizations of a data set including $40$ pulsars, with CURN parameters and timing baseline matching with either the 15-year or $12.5$-year dataset. This demonstrates that having a longer timing baseline does reduce the variance in the SNR and provides some increase in SNR value on the timescales of these datasets (5-15 years). However, the difference in the CURN amplitude between these two datasets also has a significant impact on the SNR. The larger CURN amplitude found in the $15$-year data set reduces the SNR by approximately $25\%$ over the course of a long observation baseline. 

This simple test gives us insight into why the upper limits presented in this paper are worse at some epochs. The increased amplitude of the red noise background in our data reduces the significance of GW memory signals. We plan to perform a more extensive evaluation of the GW memory signal scaling in PTA data in the future, including an update to the detection prospects work presented in \cite{islo_prospects_2019}.

\section{\label{sec:Conclusion} Conclusion} 

We performed a search for GW memory events in NANOGrav's 15-year dataset, and found no significant evidence in favor of such events. The only discernible event occurred near the end of the dataset, around an MJD of $58200$ as shown in \autoref{fig:t0_posteriors}. For this event, model selection yields an insignificant Bayes factor of 3.1 in favor of the model with GW memory. As this event occurs late in the dataset where there is not enough data remaining to well constrain GW memory build up, we believe this to be a spurious event. We will need future datasets to definitively rule this event out.

This could be due to the fact that there is not enough data remaining to build up the GW memory significance after the time the event occurs. Future datasets will be able to rule if this event is spurious.

Having found no significant events, we proceed to set new upper limits on GW memory. Overall, in the early times of sky averaged upper limits (\autoref{fig:epoch_UL}), the earth term and high strain pulsar term event rate upper limits (\autoref{fig:rate_UL}), and the epoch averaged upper limit sky maps (\autoref{fig:Sky_UL}), we see improvements in these upper limits over the previous $12.5$-year search. However, these improvements are not uniform, and the strain upper limits in central times of the 15-year data set are less constraining. We also see these constraints weaken for low-strain pulsar term event rate upper limits. This loss of sensitivity is due to the increased CURN amplitude recovered in the $15$-year dataset compared with the $12.5$-year dataset, as well as the effects of quadratic subtraction \cite{cordes_detecting_2012} compounded over the various pulsar timing baselines, which have changed between the datasets.  

While we do not expect our new constraints for the SMBHB population to be astrophysically interesting (see Fig. \ref{fig:rate_UL}), our limits apply to all astrophysical and cosmological sources of GW memory. Because all GW-emitting events act as sources for GW memory, these searches will continue to be an important catch-all for constraining more exotic GW sources.


\section{\label{sec:Acknowledgements} Acknowledgements} 

\textit{Author Contributions:} We present the author list in alphabetical order in recognition that this paper is the result of the decades long work of many people in the NANOGrav collaboration. All authors contributed to the activities of the NANOGrav collaboration leading to the work presented here, and reviewed the manuscript, text, and figures prior to the paper submission. Specific author contributions are as follows. 
G.A., A.A., A.M.A., Z.A., P.T.B., P.R.B., H.T.C., K.C., M.E.D., P.B.D., T.D., E.C.F., W.F., E.F., G.E.F., N.G., P.A.G., J.G., D.C.G., J.S.H., R.J.J., M.L.J., D.L.K., M.K., M.T.L., D.R.L., J.L., R.S.L., A.M., M.A.M., N.M., B.W.M., C.N., D.J.N., P.N., T.T.P., B.B.P.P., N.S.P., H.A.R., S.M.R., P.S.R., A.S., C.S., B.J.S., I.H.S., K.S., A.S., J.K.S., and H.M.W. through a combination of observations, arrival time calculations, data checks and refinements, and timing model development and analysis; additional specific contributions to the data set are summarized in \cite{agazie_nanograv_2023-3}.
R.B. ran the Bayesian detection and upper limit analysis, and wrote the bulk of the paper. J.S. assisted in paper writing, analysis, code debugging, and ran simulations for SNR scaling tests. B.B., J.S.H., J.T., X.S. provided guidance on searches and analysis. 

\textit{Acknowledgments:} 

The work contained herein has been carried out by the NANOGrav collaboration, which receives support from the National Science Foundation (NSF) Physics Frontier Center award numbers 1430284 and 2020265, the Gordon and Betty Moore Foundation, NSF AccelNet award number 2114721, an NSERC Discovery Grant, and CIFAR. The Arecibo Observatory is a facility of the NSF operated under cooperative agreement (AST-1744119) by the University of Central Florida (UCF) in alliance with Universidad Ana G. M$\acute{\text{e}}$ndez (UAGM) and Yang Enterprises (YEI), Inc. The Green Bank Observatory is a facility of the NSF operated under cooperative agreement by Associated Universities, Inc.
L.B.\ acknowledges support from the National Science Foundation under award AST-1909933 and from the Research Corporation for Science Advancement under Cottrell Scholar Award No.\ 27553.
P.R.B.\ is supported by the Science and Technology Facilities Council, grant number ST/W000946/1.
S.B.\ gratefully acknowledges the support of a Sloan Fellowship, and the support of NSF under award \#1815664.
The work of R.B., N.La., X.S., J.P.S., J.T., and D.W.\ is partly supported by the George and Hannah Bolinger Memorial Fund in the College of Science at Oregon State University.
M.C., P.P., and S.R.T.\ acknowledge support from NSF AST-2007993.
M.C.\ was supported by the Vanderbilt Initiative in Data Intensive Astrophysics (VIDA) Fellowship.
Support for this work was provided by the NSF through the Grote Reber Fellowship Program administered by Associated Universities, Inc./National Radio Astronomy Observatory.
Pulsar research at UBC is supported by an NSERC Discovery Grant and by CIFAR.
K.C.\ is supported by a UBC Four Year Fellowship (6456).
M.E.D.\ acknowledges support from the Naval Research Laboratory by NASA under contract S-15633Y.
TD and MTL were supported by an NSF Astronomy and Astrophysics Grant (AAG) award number 2009468 during this work. 
E.C.F.\ is supported by NASA under award number 80GSFC24M0006.
G.E.F., S.C.S., and S.J.V.\ are supported by NSF award PHY-2011772.
K.A.G.\ and S.R.T.\ acknowledge support from an NSF CAREER award \#2146016.
A.D.J.\ and M.V.\ acknowledge support from the Caltech and Jet Propulsion Laboratory President's and Director's Research and Development Fund.
A.D.J.\ acknowledges support from the Sloan Foundation.
N.La.\ acknowledges the support from Larry W. Martin and Joyce B. O'Neill Endowed Fellowship in the College of Science at Oregon State University.
Part of this research was carried out at the Jet Propulsion Laboratory, California Institute of Technology, under a contract with the National Aeronautics and Space Administration (80NM0018D0004).
D.R.L.\ and M.A.M.\ are supported by NSF \#1458952.
M.A.M.\ is supported by NSF \#2009425.
C.M.F.M.\ was supported in part by the National Science Foundation under Grants No.\ NSF PHY-1748958 and AST-2106552.
A.Mi.\ is supported by the Deutsche Forschungsgemeinschaft under Germany's Excellence Strategy - EXC 2121 Quantum Universe - 390833306.
The Dunlap Institute is funded by an endowment established by the David Dunlap family and the University of Toronto.
P.N.\ acknowledges support from the BHI, funded by grants from the John Templeton Foundation and the Gordon and Betty Moore Foundation
K.D.O.\ was supported in part by NSF Grant No.\ 2207267.
T.T.P.\ acknowledges support from the Extragalactic Astrophysics Research Group at E\"{o}tv\"{o}s Lor\'{a}nd University, funded by the E\"{o}tv\"{o}s Lor\'{a}nd Research Network (ELKH), which was used during the development of this research.
H.A.R.\ is supported by NSF Partnerships for Research and Education in Physics (PREP) award No.\ 2216793.
S.M.R.\ and I.H.S.\ are CIFAR Fellows.
Portions of this work performed at NRL were supported by ONR 6.1 basic research funding.
J.D.R.\ also acknowledges support from start-up funds from Texas Tech University.
J.S.\ is supported by an NSF Astronomy and Astrophysics Postdoctoral Fellowship under award AST-2202388, and acknowledges previous support by the NSF under award 1847938.
C.U.\ acknowledges support from BGU (Kreitman fellowship), and the Council for Higher Education and Israel Academy of Sciences and Humanities (Excellence fellowship).
C.A.W.\ acknowledges support from CIERA, the Adler Planetarium, and the Brinson Foundation through a CIERA-Adler postdoctoral fellowship.
O.Y.\ is supported by the National Science Foundation Graduate Research Fellowship under Grant No.\ DGE-2139292.

\bibliography{NG_15_BWM_bib}{}
\bibliographystyle{unsrt_et_al_15-3author_notitle}

\end{document}